\documentclass[prl, amsfonts, twocolumn, nofootinbib, showpacs]{revtex4}
\usepackage{graphicx, epsfig}
\tighten

\newcommand{\be}{\begin{equation}}
\newcommand{\ee}{\end{equation}}
\newcommand{\bea}{\begin{eqnarray}}
\newcommand{\eea}{\end{eqnarray}}
\newcommand{\B}{{\rm B}}
\newcommand{\Lep}{{\rm L}}

\def\bbox{{\,\lower0.9pt\vbox{\hrule \hbox{\vrule height 0.2 cm
\hskip 0.2 cm
\vrule  height 0.2 cm}\hrule}\,}}
\begin{document}
\title{Information-preserving black holes still do not preserve baryon
number and other effective global quantum numbers \footnote{Submitted for 
the Gravity Essay Competition 2005, Honorable Mention received.}}

\author{Dejan Stojkovic$^1$\footnote{present email: dbs3@case.edu}}
\author{Glenn D. Starkman$^{2}$}
\author{Fred C. Adams$^1$}
\affiliation{$^1$MCTP, Department of Physics, University of Michigan,  Ann
Arbor, MI 48109-1120 USA}
\affiliation{$^2$ Department of Physics, Case Western Reserve University,
             Cleveland, OH~~44106-7079}

\begin{abstract}
It has been claimed  recently that the black-hole information-loss
paradox has been resolved: the evolution of quantum states in the
presence of a black hole is unitary and information preserving. We
point out that, contrary to some claims in literature,
information-preserving black holes still violate baryon number and
any other quantum number which follows from an effective (and thus
approximate) or anomalous
 symmetry.
\end{abstract}

\maketitle

The Standard Model Lagrangian possesses several global $\rm{U}(1)$
symmetries. Each such symmetry has an associated  conserved
quantum number. Important among these are baryon number $\B$ and
lepton number $\Lep$. The baryon number of a particle is $\B=1$ if
it is a baryon, $\B=-1$ if it is an antibaryon. Consequently,
quarks carry $\B=1/3$, and  anti-quarks $\B=-1/3$, while all other
particles (e.g. leptons, gauge bosons, gravitons etc.) are
assigned $\B = 0$. Similarly, the lepton number  of every lepton
(electrons, muons, tau-particles and their associated neutrinos)
is $\Lep=1$, while the lepton number of every anti-lepton is $-1$,
and all other particles are assigned lepton number $\Lep=0$.

Since protons are the lightest baryons, baryon number conservation
would be violated if a proton could decay. Because electrons and
neutrinos are the only free fermions lighter than  protons (quarks
being confined in mesons or baryons), lepton number would also be
violated in such decays.

Processes like \be p \rightarrow e^+ + \gamma \ \ {\rm or} \ \ p
\rightarrow e^+ + \pi^0 \ee do not violate the conservation laws
of energy, electromagnetic charge, linear or angular momentum.
However, they do not occur in nature (at least at the energies so
far probed) because they violate the conservation of baryon and
lepton number. The apparent stability of the proton and the lack
of other similar baryon or lepton number violating processes are
both consequences and manifestations of the conservation of baryon
and lepton numbers.

Despite the important role of B and L conservation in low-energy
physics, it is widely believed that these are both violated at
higher energy. Within the Standard ($\rm{SU}(2)\times\rm{U}(1)$)
Model of electroweak
 interactions, both B and L are anomalous symmetries, conserved by
perturbative processes (in the coupling constants $\alpha_1$ and
$\alpha_2$) but violated by non-perturbative processes such as
those
 mediated by
instantons. At temperatures or energies well below the
electro-weak  symmetry-breaking scale ($v_{EW} \simeq 250$GeV),
these non-perturbative processes are suppressed by the extremely
small factor $e^{-8\pi/\alpha}$ (giving, for example, a proton
life-time of $10^{141}$yr  \cite{AL}). Above this scale, B and L
violation are expected to be essentially  unsuppressed, although
$\B-\Lep$ remains conserved within the Standard Model.

Similarly, in the context of any grand unified theory (GUT),
baryon-number and lepton-number violating processes would be
expected to be generic since the quarks and leptons of a given
family would be members of the same representation of the GUT
 gauge group.  The GUT gauge bosons would therefore mediate
transformations between quarks and leptons, just as SU(2) gauge
bosons mediate transformations between up and down type quarks or
between charged leptons and their associated neutrinos.

Thus, while neither electroweak instanton mediated processes nor
GUT lepto-quark boson mediated processes  have yet been observed,
B and L violating processes, such as proton decay, are expected to
occur at low energies, albeit with extremely low probability.
Moreover, there {\bf is} strong circumstantial evidence for B
violation -- the overwhelming predominance of baryons over
anti-baryons in the universe. Non-conservation of baryon number is
one of the three key ingredients in most models of baryogenesis
\cite{Sakharov}.

We thus see that B and L  are approximately, albeit nearly
perfectly, conserved in the low-energy effective theory of the
Standard Model.

One arena in which global charge violation has been expected to
occur is inside black holes. Formation and subsequent evaporation
of a black hole may lead to the so-called information loss paradox
-- an initially pure state can evolve into a mixed state, thus
violating quantum coherence. Often in the literature, this
information loss is connected with non-conservation of baryon
number. Black holes can carry local gauge charges, but cannot
carry global charges such as B or L. Since information can not be
pulled back once it crosses the horizon, it is usually said that
the baryon number of the initial state of a black hole precursor,
or of any material thrown into a black hole,  is "forgotten" by
the black hole. The baryon number of the final state is thus
independent of the baryon number of the initial state. Thus, it is
said, information loss implies baryon number non-conservation.

A direct consequence of a black hole's erasure of the baryon
number of an initial incoming state, is that particle processes
mediated by virtual black holes can be baryon number violating. On
short distances (at high energies), fluctuations in the space-time
metric are expected to be large. These fluctuations can be
described effectively as virtual black holes \cite{Hawking} and
exemplify the possibility of quantum gravity mediated baryon
number violation. Since they are quantum gravitational, one
expects these processes to be suppressed by inverse powers of the
energy scale associated with quantum gravity, $M_F$.

The quantum gravity energy scale is normally taken to be $M_F \sim
M_{Pl} \simeq 10^{19}$GeV. This is very high compared to any other
scale in nature, so the  probability for quantum gravitational
processes at low energy will be extremely small. The only exception
might be processes involving fundamental scalar fields (Higgs,
axions, quintessence field,  etc.) where one can imagine processes
that are not highly suppressed. The other context where such
processes can be problematic is models where the quantum gravity
energy scale is much lower than $M_{Pl}$, for example brane world
models where typically $M_F \sim 1-10$~TeV. Proton decay and similar
processes, if directly suppressed only by inverse powers of $M_F$,
would be catastrophic in these models \cite{AKP}, and require rather
elaborate fixes, such as placing quarks on one brane and leptons on
another\footnote{This in turn can radically modify some of the basic
predictions of the model \cite{SS}}.

The debate whether black holes preserve information has been
recently renewed after Hawking's claim that the information loss
paradox has been resolved in favor of information preservation.
(Since there is no scientific publication available yet, we refer
to \cite{Dublin}.) The basis of the claim is that black hole
formation and subsequent evaporation can be thought of as a
scattering process. One sends in particles and radiation from
infinity and measures what comes out at infinity. One never probes
what happens in the intermediate region. On short time scales
(shorter than the life-time of the black hole) one might observe
events that appear as information loss. On times scales longer
than the life-time of the black hole, the black hole evolution
must be unitary and the information is preserved. If one observes
the initial state (with no black hole, since it has not been
formed yet) and the final state (with no black hole, since it has
evaporated), then the evolution must be unitary. Non-trivial
contributions from the black hole state decay exponentially with
time and do not contribute in the limit where one observes the
final state at temporal infinity. This gives a unitary mapping
from the initial state to the final state.

In literature, one can often find cliams that if black holes
preserve information, then baryon and lepton number violating
processes cannot be mediated by virtual black holes. Arguments of
this type have been used, for example, to question the limits on
TeV-scale gravity \cite{K}. Contrary to this point of view, we
argue that information preserving black holes still violate baryon
number. Our conclusion rests on the observation made above that
baryon number is only a low-energy effective quantum number even
in the absence of quantum gravity. Viewed from an information
theoretic perspective, it is important to realize that when a
proton decays, through for example lepto-quark boson mediation in
a GUT, no information is lost, any more than information is lost
when a neutron decays into a proton by emitting an (off-shell)
W-boson (which then ``decays'' into an electron and
electron-anti-neutrino). Similarly, electroweak instanton mediated
B-violating processes do not destroy information.  They conserve
all the exact quantum numbers and conserved quantities of the
initial state:  energy, momentum, total angular momentum, electric
charge, $\B-\Lep$, {\it etc}..

How do B and L get erased inside a black hole? Inside the horizon,
the in-falling matter collapses and is compressed. When it reaches
GUT-scale densities, $\rho \sim 10^{78} g/cm^3$, the system almost
immediately becomes neutral with respect to baryon charge regardless
of its initial value (see \cite{Frolov} and references therein).
This depends on the existence of a GUT. However, one expects that
far sooner, the density of the collapsing matter will be
characteristic of the electroweak scale, $\rho \sim 10^{27}g/cm^3$,
and $\B$ and $\Lep$ will be erased by ``over-the-barrier"
non-perturbative electroweak processes. Indeed, any new
beyond-the-standard-model physics for which $U(1)_\B$ is just a
low-energy effective global symmetry will produce the same effect
long before the matter reaches the Planck density, $\rho_{Pl} \sim
10^{94} g/cm^3$. Finally, when matter reaches the Planck density,
quantum gravity mechanisms become important (e.g. wormholes). These
are also capable of erasing any initial baryon number before the
infalling matter crosses the singularity.

This mechanism for destruction of baryon charge should be
effective for macroscopic black holes, which live for long time.
For virtual black holes there is no definite answer without fully
understood quantum gravity. However, there are strong indications
that the conclusion remains similar. The rate of global charge
disappearance inside horizon was calculated in \cite{Coleman} and
shown to be exponential. Thus, even a virtual black hole that
lives only one Planck time would get rid of at least $e^{-1}$ of
the original baryon charge. [In \cite{Pawl1}, the timescale for a
related effect --- loss of massive vector hair by a black hole ---
was analyzed. When applied appropriately, calculations are in
agreement with \cite{Coleman}.]

According to these arguments,  even an information preserving
black hole can erase the baryon number of the initial state. This
is not a paradox since baryon number is not actually conserved, it
is only approximately conserved at low energy, temperature and
density.

We can treat  other global charges on equal footing with baryon
charge. In principle, all the charge conservation laws that follow
from the effective global symmetries or low energy approximative
symmetries (like baryon number B, lepton number L, individual
generational lepton number $\Lep_i$, charge conjugation C, parity
P, the combined symmetry $CP$, {\it etc.}) can be violated by
black holes whose evolution is unitary (information preserving).
Global charges whose conservation  might not be violated include
$\B-\Lep$, since  $U(1)_{\B-\Lep}$ is preserved by the Standard
Model and by many possible GUTs, and CPT,  because no
self-consistent local field theory accommodates CPT violation.
However, since there is no dynamics that protects these
symmetries, it is possible that they too can be violated. Finally,
charges whose conservation can not be violated are those that
follow from unbroken gauge symmetries (continuous like $U(1)_{EM}$
or discrete), and space-time symmetries (energy, momentum and
angular momentum).

Charges protected by topological reasons (domain wall kink number,
string and monopole winding numbers, {\it etc.}) deserve special
attention.

Point-like defects like magnetic monopoles can be swallowed by a
black hole. Since magnetic monopoles are charged with magnetic
gauge charge, a black hole cannot violate a net monopole number
conservation. If a monopole-antimonopole pair is swallowed by a
black hole, it can annihilate (unwind) inside \cite{SF} without
violating a net topological charge. If a number of monopoles of
the same magnetic charge is swallowed, the black hole becomes
magnetically charged and cannot evaporate completely. Consider a
black hole that captures $N \gg 1$ monopoles. A monopole is a
highly coherent state of many gauge quanta and emission of a
monopole by a black hole is  highly suppressed. Even if this
 process
is somehow allowed, the Hawking temperature of a black hole becomes
of order of a monopole mass only at the end of evaporation, and a
black hole could radiate only a few monopoles. Generically, since
the Hawking radiation can not violate the gauge symmetry, a black
hole can not evaporate completely. Instead it leaves a remnant ---
an extreme magnetically charged Reissner-Nordstrom black hole. The
mass of the  (non-rotating) remnant $M_r$ must be greater than the
magnetic charge $Q_m$ of the black holes, $M_r \geq Q_m$, or
otherwise the remnant would be a naked singularity. The Hawking
temperature of an extremal black hole is zero and such a black hole
does not evaporate further. Thus, the net topological charge
 remains conserved.

The situation with extended defects is more complicated. It was
shown in \cite{FSS} that there is a non-vanishing energy and
angular momentum flux through the black-hole horizon in
non-stationary black hole-string and black hole-domain wall
 configurations.
This implies that, in principle, a black hole can accrete energy
from an extended topological defect. In the case of a finite size
defect, i.e., a cosmic string ending on monopoles or a domain wall
bounded by a string, the final configuration would be a black hole
with a defect swallowed within a horizon. If the defect did not
carry any gauge charges, such a black hole could evaporate
completely. In this process, the total topological charge would not
be violated since finite size defects (cosmic strings ending on
monopoles or domain walls bounded by strings) have a trivial net
topology. Infinite strings and domain walls are configurations with
a non-trivial topology. However, they can not be completely
swallowed by a black hole since the part of the defect that is
accreted within a horizon gets replaced with a part pulled out from
infinity.

The other question related to the previous discussion is nucleation
of black holes within a defect. It is well known that a cosmic
string can break if a monopole-antimonopole pair is nucleated on the
string. In a similar way a cosmic string can break if a pair of
black holes is nucleated on it. The analog process in a domain wall
case would be a black string loop (a one dimensional generalization
of a black hole solution) nucleating on a domain wall world sheet
\cite{SFS} \footnote{Note that there is no known solution of such a
configuration. A simple real scalar field with broken $Z_2$ symmetry
does not allow for such solutions but some more complicated models
may in principle.}. A finite defect would be broken into a finite
number of pieces by these processes. However, a finite defect has a
trivial net topology. An infinite defect with non-trivial topology
can never be broken into a finite number of pieces in finite time.
We thus conclude that true topological charges can not be broken by
processes induced by black holes. We note, though, that in practice
we often meet objects that do not meet the strict criteria of a
truly topological configuration. For example cosmic strings and
domain walls that could arise in phase transitions in early
universe, though possibly larger than a horizon size, are still
finite objects.

To summarize, a black hole whose evolution preserves information
in the strict sense makes distinction between different types of
quantum numbers:
\newline\noindent
1. quantum numbers that are violated:\newline\noindent
  {\it eg.}  $\B$, $ \Lep$, $ \Lep_i$, C, and CP;\newline
\noindent 2.  quantum numbers that {\it might } not be
violated:\newline\noindent {\it eg.}  $ {\B-\Lep }$, and CPT;
\newline \noindent 3. quantum numbers that are not violated within
the currently understood theory:\newline\noindent {\it eg.}
$Q_{EM}$, E, ${\vec L}$, ${\vec p}$, and true  topological
 charges.

This distinction is in agreement with an argument given by Hawking
in \cite{Hawking}. There, probability amplitudes for processes
mediated by virtual black holes were derived using Euclidian path
integral formalism. After averaging over all diffeomorphisms (all
metrics that contribute to a particular process), the formalism
gives a zero transition amplitude unless energy is
 conserved.
After averaging over all the gauge degrees of freedom, the
formalism gives a zero transition amplitude unless the gauge
charges
 are conserved.
Since one does not average over global symmetries, a transition
amplitude can be non-zero even if global charges are not
 conserved.
The conclusion that is drawn in \cite{Hawking} is that processes
where quantum coherence is lost may lead to non-conservation of
 global charges.
The point we are making in this paper is that even black holes
whose evolution preserves the information do not conserve certain
global
 charges, such as baryon and lepton number.
This is not in contradiction with conclusions in \cite{Hawking},
since  baryon number non-conservation does not imply information
loss.

Arguably, the most important consequence of this statement is that
virtual black holes can mediate proton decay in models with low
energy scale quantum gravity. Specifically, the claim that black
holes preserve information does not prevent proton decay of this
type. There may be other ways to protect the proton like gauging
the baryon number
 \cite{Pawl2} or other more exotic mechanisms \cite{Nima}; however, to the
extent that these are successful, it seems that they must suppress
baryon-number violation at all energy  scales below the quantum
gravity scale, and hence at all times in the history of  the
universe. In order for them to be compatible with generic models
of baryogensis \cite{Sakharov} some modifications that allow for
baryon number violation in the early universe are necessary.

\vspace{12pt} {\bf Acknowledgments}:\ \ The authors are grateful
to Malcolm Perry, Don Page, James Wells and Gordon Kane for very
useful
 conversations.
The work was supported by the DOE grants to the Michigan Center
for Theoretical Physics, University of Michigan, and to the
particle astrophysics group at Case Western Reserve University.
GDS acknowledges the hospitality of the MCTP where this work was
done.

\vspace{12pt} {\bf Note added}:\ \ After the completion of this
work, a long awaited paper by S. Hawking appeared on the web
\cite{*}. At the end of the paper, the author also addresses the
question of baryon number violation by information preserving black
holes. The author's conclusion differs from the one presented here.
We leave the judgment to the reader.


\begin{thebibliography}{9}

\bibitem{AL} F. C. Adams, G. Laughlin,  Rev. Mod. Phys. {\bf 69},
337 (1997)

\bibitem{Sakharov} A. Sakharov, Pisma Zh.Eksp.Teor.Fiz. {\bf 5} 32 (1967),
JETP Lett. {\bf 5}24 (1967); Sov.Phys.Usp. {\bf 34} 392 (1991),
Usp.Fiz.Nauk {\bf 161} 61 (1991)

\bibitem{Hawking} S. Hawking,  Phys. Rev. {\bf D53}, 3099 (1996)

\bibitem{AKP} F.C. Adams, G.L. Kane, M. Mbonye and M.J. Perry,
Int. J. Mod. Phys. {\bf A16}, 2399 (2001)

\bibitem{SS} D. Stojkovic, Glenn D. Starkman, D.C. Dai,
 Phys. Rev. Lett. {\bf 96} 041303 (2006)


\bibitem{Dublin} S. Hawking talk given at GR17,
the 17th International Conference on General Relativity and
Gravitation, 18th - 23rd July 2004, Dublin, Ireland

\bibitem{K} A. B. Kobakhidze, Phys. Lett. {\bf B514} 131 {2001}

\bibitem{Frolov} V. Frolov and I. Novikov. {\em Black Hole Physics: Basic
Concepts and New Developments} (Kluwer Academic Publ.), 1998.

\bibitem{Coleman} S. R. Coleman, S. Hughes, Phys. Lett. {\bf B309} 246
 {1993}

\bibitem{Pawl1} A. Pawl, Nucl. Phys. {\bf B679} 231 (2004)

\bibitem{SF} D. Stojkovic,  K. Freese, Phys.  Lett. {\bf B606}, 251
(2005)

\bibitem{FSS} V. P. Frolov, D. V. Fursaev, D. Stojkovic, Class.
Quant. Grav. {\bf 21} 3483 (2004); JHEP {\bf 0406} 057 (2004); D.
Stojkovic,  Phys. Rev. Lett  {\bf 94}, 011603 (2005);JHEP {\bf
0409} 061 (2004); V. P. Frolov, M. Snajdr, D. Stojkovic, Phys.
Rev. {\bf D68} 044002 (2003)

\bibitem{Pawl2} L. M. Krauss and F. Wilczek, Phys. Rev. {\bf D62} 1221
(1989);  A. Pawl,  hep-th/0501005

\bibitem{Nima} N. Arkani-Hamed, M. Schmaltz, Phys. Rev. {\bf D61} 033005
 (2000)

\bibitem{SFS} D. Stojkovic, K. Freese, G.D. Starkman, Phys. Rev. {\bf D72} 045012 (2005)

\bibitem{*} S.W. Hawking, hep-th/0507171

\end{thebibliography}
\end{document}